\documentclass[conference]{IEEEtran}

\makeatletter
 \let\old@ps@headings\ps@headings
 \let\old@ps@IEEEtitlepagestyle\ps@IEEEtitlepagestyle
 \def\confheader#1{%
 \def\ps@headings{%
 \old@ps@headings%
 \def\@oddhead{\strut\hfill#1\hfill\strut}%
 \def\@evenhead{\strut\hfill#1\hfill\strut}%
 }%
 \def\ps@IEEEtitlepagestyle{%
 \old@ps@IEEEtitlepagestyle%
 \def\@oddhead{\strut\hfill#1\hfill\strut}%
 \def\@evenhead{\strut\hfill#1\hfill\strut}%
 }%
 \ps@headings%
 }
\makeatother

\IEEEoverridecommandlockouts

\usepackage{cite}
\usepackage{amsmath,amssymb,amsfonts}
\usepackage{algorithmic}
\usepackage{graphicx}
\usepackage{todonotes}
\usepackage{textcomp}
\usepackage{xcolor}
\usepackage[T1]{fontenc}
\def\BibTeX{{\rm B\kern-.05em{\sc i\kern-.025em b}\kern-.08em
    T\kern-.1667em\lower.7ex\hbox{E}\kern-.125emX}}
\begin{document}
\title{Ultra-Precise Synchronization for TDoA-based Localization Using Signals of Opportunity}

\author{
\IEEEauthorblockN{Thomas Maul, Joerg Robert}
\IEEEauthorblockA{\textit{Technische Universität Ilmenau} \\
\textit{M2M Research Group}\\
Ilmenau, Germany \\
\{thomas.maul, joerg.robert\}@tu-ilmenau.de}
\and
\IEEEauthorblockN{Sebastian Klob}
\IEEEauthorblockA{\textit{Friedrich-Alexander Universität Erlangen-Nürnberg (FAU)} \\
\textit{Information Technology (Communication Electronics)}\\
Erlangen, Germany \\
sebastian.klob@fau.de}}
\IEEEaftertitletext{This work has been submitted to the IEEE for possible publication. Copyright may be transferred without notice, after which this version may no longer be accessible.}
\maketitle

\begin{abstract}
Precise localization is one key element of the Internet of Things (IoT). Especially concepts for position estimation when Global Navigation Satellite Systems (GNSS) are unavailable have moved into the focus. One crucial component for localization systems in general and precise runtime-based positioning, in particular, is the necessity of ultra-precise clock synchronization between the receiving base stations. Our work presents a software-based approach for the wireless synchronization of spatially separated base stations using a low-cost off-the-shelf frontend architecture. The proposed system estimates the time synchronization, sampling clock offset, and carrier frequency offset using broadcast signals as Signals of Opportunity.
\\In this paper, we derive the theoretical lower bound for the estimation variance according to the Modified Cramer-Rao Bound. We show that a theoretical time synchronization accuracy in the range of ps and a frequency synchronization precision in the range of milli-Hertz is achievable. An algorithm is presented that estimates the desired parameter based on evaluating the Cross-Correlation Function between base stations. Initial measurements are conducted in a real-world environment. It is shown that the presented estimator nearly reaches the theoretical bound within a time and frequency synchronization accuracy of down to 200 ps and 6 mHz, respectively.
\end{abstract}


\renewcommand{\baselinestretch}{0.965}\normalsize
\section{Introduction}
Precise localization is one key element of the Internet of Things (IoT). The application areas are very diverse, ranging from asset tracking in industrial applications to navigation in autonomous driving. In classical outdoor environments, well-known Global Navigation Satellite Systems (GNSS) like GPS or Galileo are the most often deployed systems due to global coverage and good accuracy. Nevertheless, these systems exhibit severe drawbacks, e.g., receiver costs or insufficient accuracy in indoor scenarios \cite{GPS_Indoor}. There exist several systems with different approaches for indoor localization. A short overview is given in \cite{IndoorLocOverview}, where it is noticeable that the most promising solutions in terms of accuracy are runtime-based Ultra Wideband technology (UWB) systems, as stated in \cite{UWBTDoA}. 
\\A necessity in runtime-based localization is the ultra-precise clock synchronization of the receiving base stations, namely $3\,\text{ns}$, to achieve sub-meter accuracy. 
A variety of methods has already been investigated in this context. A cable-based synchronization approach is often proposed in indoor environments, offering excellent accuracy in comparison to wireless systems \cite{Wired_Wireless_Comparison}. Major drawbacks are the high installation costs and the infeasibility when used in spatially separated locations. Another approach is the usage of software-defined time references like the Precise Time Protocol (PTP) \cite{PTP}. Despite the ease of implementation, this approach is not considered in our work because the expected precision is in the range of several hundreds of ns and therefore does not allow sub-meter accuracy. GNSS could also be used for the synchronization of base stations. There exist ultra-precise GNSS-based modules for clock synchronization, e.g., the ublox NEO-F10T1\footnote{https://www.u-blox.com/en/product/neo-f10t-module \newline (accessed May 2023)}. The module's specified precision is 10 ns, adequate for many localization scenarios but insufficient for sub-meter accuracy. In addition, GNSS signals may not be available in indoor environments. This paper proposes a novel synchronization approach based on Signals of Opportunity (SoO), exiting the work presented in \cite{Troeger}.
Examples of SoO are broadcast or mobile communication signals, which have not been radiated for synchronization. The remainder of this paper is structured as follows:
\\ Section \ref{Concept} provides an overview of the fundamental system concept. Section \ref{Theory} derives the theoretical limits of the investigated concept. Section \ref{Framework} proposes an estimation algorithm. Section \ref{Results} validates the proposed algorithm against the fundamental limits. Finally, section \ref{Conclusion} gives a conclusion.

\section{System Concept} \label{Concept}
\begin{figure}[htbp]
\centerline{\includegraphics[width=0.5\textwidth]{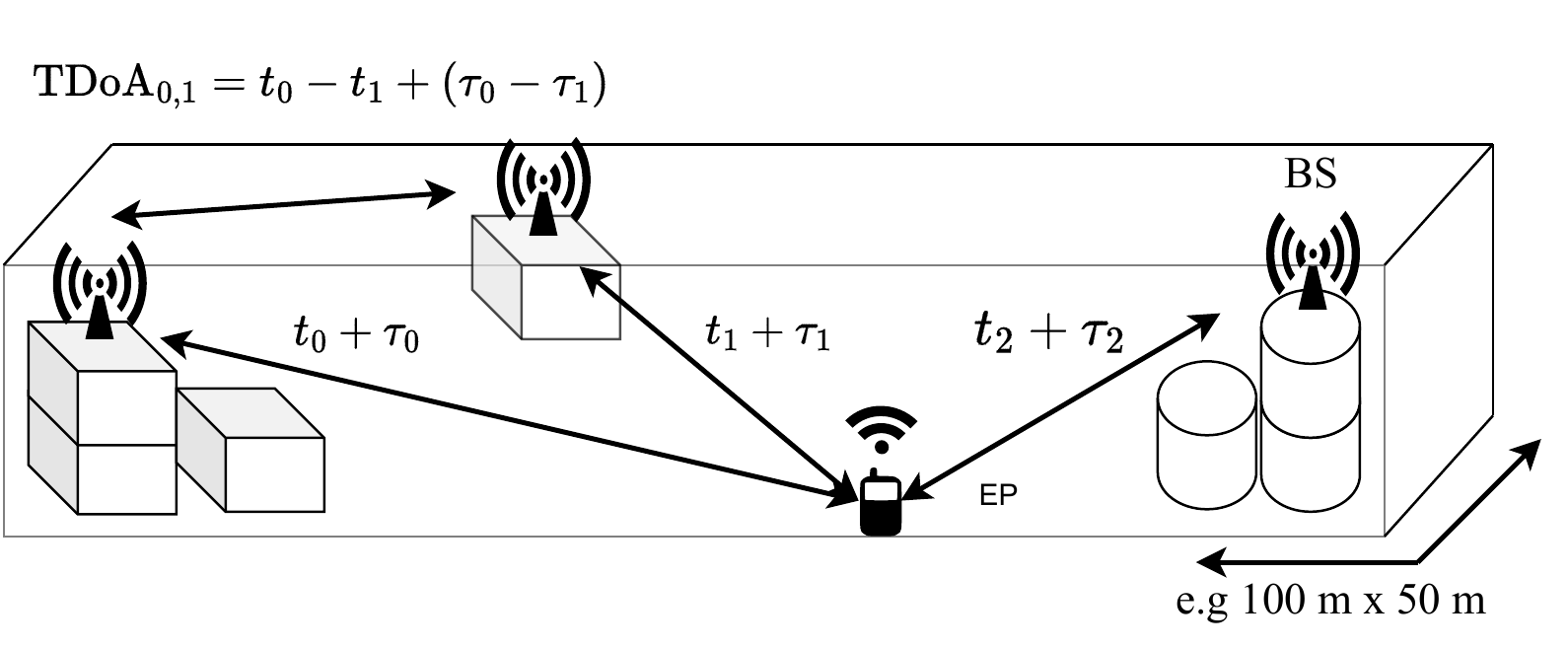}}
\caption{Considered scenario with multiple base stations (BS) in a warehouse localizing a mobile endpoint (EP) using TDoA measurements}
\label{TDoA_Concept}
\end{figure}

This section overviews the underlying system concept for synchronization using SoO. We start by considering a classical indoor localization scenario in a warehouse environment using Time-Difference-of-Arrival (TDoA) measurements, depicted in Fig. \ref{TDoA_Concept}. For a deeper understanding of the necessity for synchronization, we will first look at the principle of runtime-based localization. Multiple base stations (BS) - at least three for 2D positioning - are spatially distributed throughout the warehouse area. The object to be localized, further denoted as endpoint (EP), emits a localization waveform that reaches the base station with index $i$ within a runtime of $t_i$. At the base station, the signal is detected at time \mbox{$t_i + \tau_i$}, whereas $\tau_i$ denotes the clock offset of a base station since an arbitrarily chosen reference time \mbox{$t_\text{ref} = 0\,\text{s}$}. After taking the difference between two base stations, e.g., the base stations with index 0 and 1, we obtain the $\text{TDoA}_{0,1}$ value, that now depends both on the desired runtime difference of the electromagnetic wave \mbox{$t_0 - t_1$} but also on the unwanted clock offsets of each base station \mbox{$\tau_0 - \tau_1$}. Estimating these unwanted clock offsets is crucial in high-precision TDoA localization and is commonly called synchronization. 
\\Furthermore, this scenario illustrates the requirements for localization and synchronization accuracy. Assuming a warehouse size of \mbox{$100\,\text{m}$ x $50\,\text{m}$}, the localization accuracy has to be significantly better than the size of the area to obtain beneficial localization results. This is why we constrain the synchronization accuracy to be 100 times better than the actual size of the warehouse. Thereas follows a required sub-meter accuracy for synchronization of less than $3\,\text{ns}$.

\subsection{Frontend Architecture for Synchronization}
\begin{figure}[htbp]
\centerline{\includegraphics[width=0.5\textwidth]{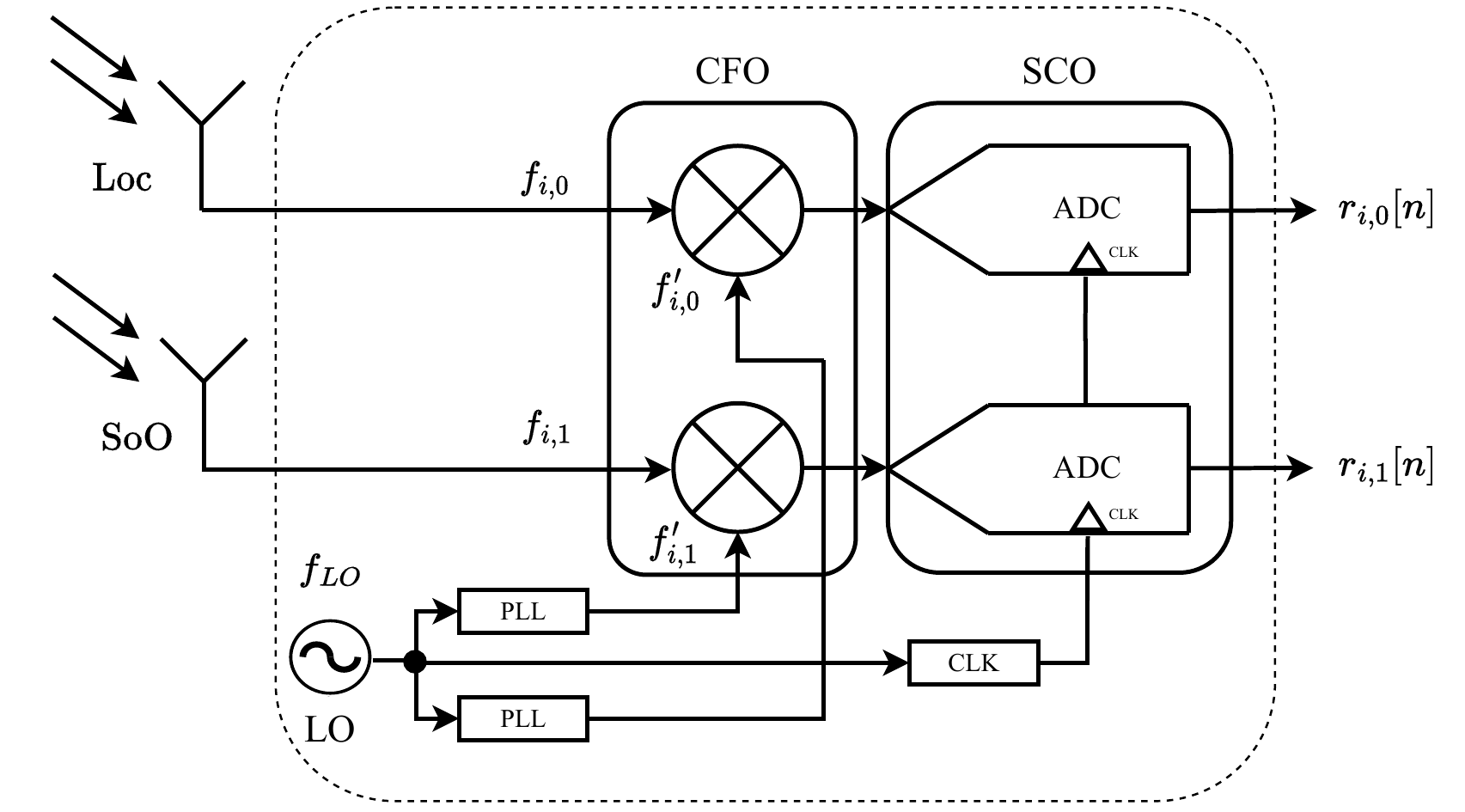}}
\caption{Base Station (BS) consisting of a dual-channel SDR frontend with a shared oscillator (LO) for the reception of the localization waveform (Loc) and the synchronization waveform (SoO)}
\label{frontend}
\end{figure}

To meet these demanding requirements, we propose a system capable of synchronizing the considered base stations from Fig. \ref{TDoA_Concept} to enable sub-meter localization accuracy in the considered warehouse scenario. Therefore, each base station is equipped with a dual-channel frontend that receives not only the localization waveform but also the SoO. This dual-channel frontend is making use of a hardware architecture that is called LO-sharing. This principle is embedded in many state-of-the-art Software Defined Radio (SDR) frontends, like the well-known Ettus TwinRX Daughterboard\footnote{https://kb.ettus.com/TwinRX/ (accessed May 2023)}. Fig. \ref{frontend} illustrates this principle in a simplified representation, depicting only the components related to the synchronization issue.

The architecture from Fig. \ref{frontend} is based on two receiving channels with a shared oscillator (LO) that is fixed to a specific frequency $f_{LO}$. The oscillator signal is feeding two independent Phase-locked loops (PLL) for each channel, generating the desired frequency for downconversion $f_{i,j}$, where $i$ is the index of the considered frontend and $j$ the index of the channel. In addition, the oscillator is also feeding the Analog to Digital Converter (ADC), where a clock signal (CLK) is used to sample the signals after downconversion. In real-world applications, the installed oscillator is a noisy component. This manifests in phase noise in the generated output signal. A full description is presented in \cite{PN}, showing that the phase noise can be separated into multiple processes with distinguishable characteristics. For reasons of simplification, we will limit ourselves to the assumption that a noisy oscillator has a frequency error, or in other words, is not ideally reaching its expected nominal frequency. This leads to an erroneous mixing frequency $f^{\prime}_{i,j}$ in the downconversion stage of the frontend, which can be observed as carrier frequency offset (CFO), denoted as $\epsilon$. Furthermore, the noisy clock signal results in a sampling clock offset (SCO) in the ADC, represented as $\xi$. This causes an erroneous symbol duration $T^\prime$. Modeling all imperfections of the oscillator, the baseband signal, denoted as $r[n]$, can be expressed as: 
\begin{equation}
    r_{i,j}[n] = r(nT+ \xi_{i,j} + \tau_{i,j})\,\text{exp}(-j2\pi \epsilon_{i,j} n T + \phi_{i,j}) + w_{i,j}[n] 
\label{baseband}
\end{equation}
where $n$ is the sampling index, $\phi_{i,j}$ is the residual phase offset of the carrier signal, and $w_{i,j}[n]$ is thermal noise, modeled as additive white Gaussian noise process (AWGN). $\tau_{i,j}$ is the timing offset since the reference time $t_{\text{ref}}$. The parameter $\epsilon_{i,j}$ models the residual CFO as the difference between the actual receive frequency $f_{i,j}$ and the noisy frequency $f^{\prime}_{i,j}$ used for downconversion of the received signal:
\begin{equation}
    \epsilon_{i,j} = f_{i,j} - f^{\prime}_{i,j}
\end{equation}
A similar approach can be taken to characterize the SCO $\xi_{i,j}$, which is defined as the difference between $f_{i,j}$ and $f^{\prime}_{i,j}$, normalized to $f_{i,j}$:
\begin{equation}
    \xi_{i,j} = \frac{f_{i,j} -f^{\prime}_{i,j}}{f_{i,j}} = \frac{\epsilon_{i,j}}{f_{i,j}}
\label{sco}
\end{equation}

The critical component of the system architecture from Fig. \ref{frontend} is the shared oscillator between the reception channels. This results in an identical influence of the noisy oscillator on both reception channels. As a direct consequence, the described LO-sharing allows estimating the synchronization parameter $\tau$, $\epsilon$, and $\xi$ using the SoO waveform and compensation of the localization waveform with the estimated parameters. 
\\A particularity in TDoA-based localization is that synchronization is especially required between mutual base stations rather than to an absolute reference time. To synchronize two base stations mutually concerning the parameter $\tau$, $\epsilon$, and $\xi$, we now introduce differential synchronization parameters as follows:
\begin{align}
    \Delta \tau &= \tau_{0,1} - \tau_{1,1} \overset{!}{=} 0 \label{const_tau}\\
    \Delta \epsilon &= \epsilon_{0,1} - \epsilon_{1,1} \overset{!}{=} 0 \label{const_eps}\\
    \Delta \xi &= \xi_{0,1} - \xi_{1,1} \overset{!}{=} 0\label{const_xi}
\end{align}
The indices are exemplary chosen so that we want to synchronize base stations with indexes 0 and 1, whereas the SoO waveform is received in channel 1. Eq. (\ref{const_tau}) constrains that there is no timing offset between mutual base stations. Eq. (\ref{const_eps}) ensures that no residual carrier frequency offset exists. Finally, (\ref{const_xi}) forces each base station's symbol duration $T^\prime$ to be identical. 
\subsection{Choice of the SoO Waveform}
Another critical feature of the proposed concept is the choice of the SoO. The required high precision and demanded suitability for indoor scenarios are very demanding. One key feature is good coverage of the SoO in the area we want to perform the synchronization. Furthermore, the received power has to be as high as possible to achieve a high signal-to-noise ratio for best performance. Another requirement that already disqualifies many waveforms is a continuous reception of the signal to enable an also continuous availability of the synchronization. The last constraint on the signal is a large usable bandwidth because this results in the best performance in terms of synchronization accuracy. This will be examined in detail in Section \ref{Theory}. According to \cite{Troeger}, one SoO that meets all requirements is digital terrestrial broadcasting standard Digital Audio Broadcast (DAB). It allows a continuous reception over a big coverage area, at least in many European countries. The network planning also ensures a high receive power, even in indoor environments. This promises a high signal-to-noise ratio. The Orthogonal Frequency-Division Multiplexing (OFDM) waveform also features a high bandwidth of $B =  1.536\,\text{MHz}$. All the above reasons lead us to choose DAB as the SoO for the following considerations in this article.


\section{Theoretical Limits for Synchronization} \label{Theory}
This section derives the theoretical limits regarding synchronization accuracy using DAB signals as SoO. Therefore, we will evaluate the so-called Modified Cramer-Rao Lower Bound (MCRB) \cite{MCRB}. This bound gives the lowest possible estimation variance for an unbiased estimation considering the underlying waveform and the signal-to-noise ratio $E_s/N_0$. 
\subsection{MCRB for Time Synchronization}
In the first step, the MCRB for the estimation of the time synchronization parameter $\tau$ is derived, which can be found in \cite{MCRB}. A limitation arises from the assumed waveform. The bound is only valid for continuous waveforms like Phase-Shift-Keying (PSK). Nevertheless, the considered DAB signals use the more complex OFDM waveform. However, for reasons of simplicity, we will use the bound stated in \cite{MCRB}. Thereby, two assumptions are made inherently on the DAB waveform. Firstly, we suppose that all subcarriers within the total signal bandwidth can be used for parameter estimation. Additionally, we assume that each symbol, including some guard intervals, can be used for parameter estimation. If these conditions are violated, one can be concerned that the MCRB tends to be too optimistic or, in other words, too loose. However, measurements in Section \ref{Results} will prove this is not the case. 
\\ Finally we start with the derivation of the synchronization parameter $\tau$ by considering the bound from \cite{MCRB}:
\begin{equation}
    \text{MCRB}(\tau) = \frac{T^2}{8\pi^2 L \Gamma}\frac{N_0}{E_S}
\label{MCRB_TAU}
\end{equation}
Since the MCRB gives the lowest estimation variance, its unit is $s^2$. However, this paper uses $\sqrt{\text{MCRB}}$, referring to the standard deviation of the estimated parameter with unit $s$. $L$ denotes the observation duration in samples and $T$ is the symbol duration. $\Gamma$ is a scaling parameter in the denominator, given as follows \cite{MCRB}:
\begin{equation}
    \Gamma = \frac{\int_{-\infty}^\infty T^2f^2|G(f)|^2 df}{\int_{-\infty}^\infty |G(f)|^2 df}
\label{Integration}
\end{equation}
It depends on the spectrum of the considered signal. More precisely, the parameter considers the bandwidth and the shape of the spectrum. In case of using DAB signals the spectrum $G(f)$ can be assumed as a rectangle with a bandwidth of $B$ and an arbitrary amplitude of $\chi$ \cite{DAB_SPECTRUM}:
\begin{equation}
    |G(f)| = \begin{cases}
        \chi & -\frac{B}{2} < f < \frac{B}{2} \\
        0 & \, \text{else}
    \end{cases}
\end{equation}
After conducting the integration from (\ref{Integration}) we arrive at a closed-form solution of the scaling parameter $\Gamma$ as:
\begin{equation}
    \Gamma = \frac{\chi ^2 T^2B^2}{12 \chi ^2} = \frac{T^2B^2}{12}
    \label{gamma}
\end{equation}
Whereas $\Gamma$ depends only on the bandwidth $B$ and the symbol duration $T$. We see that the actual amplitude $\chi$ of the spectrum cancels out. Using (\ref{gamma}) in (\ref{MCRB_TAU}) yields:
\begin{equation}
    \text{MCRB}(\tau) = \frac{3}{2\pi^2LB^2}\frac{N_0}{E_S}
\label{MCRB_TAU_CLOSED}
\end{equation}
It follows that the MCRB depends on the observation length $L$ and the bandwidth $B$. Increasing $L$ generates linearly more precise estimates of the synchronization parameter $\tau$. However, the main influence parameter is the bandwidth of the used signal. From (\ref{MCRB_TAU_CLOSED}), we see that doubling the bandwidth increases the estimation precision by a factor of 4. 
\begin{figure}[htbp]
\centerline{\includegraphics[width=0.5\textwidth]{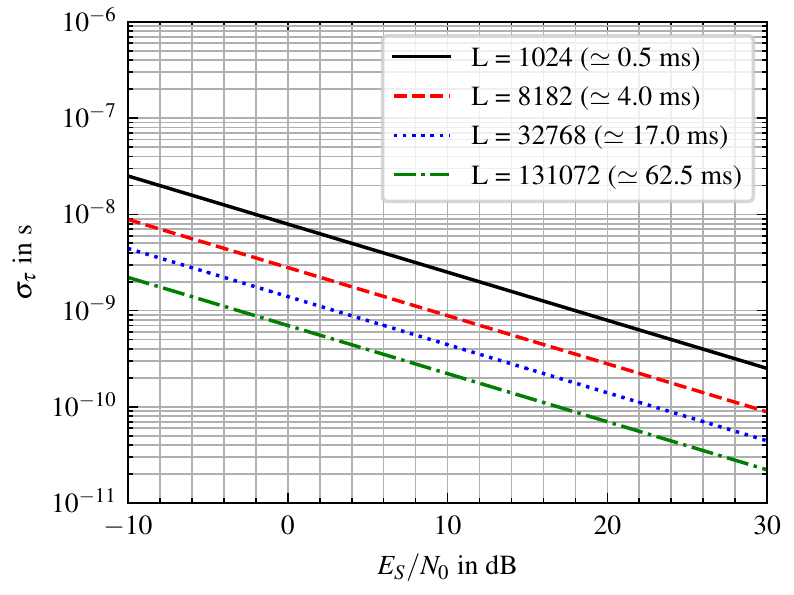}}
\caption{$\sqrt{\text{MCRB}(\tau)}$ for estimation of the time synchronization as a function of the signal-to-noise ratio $E_S/N_0$ for different observation lengths $L$}
\label{DAB_TAU}
\end{figure}
Fig. \ref{DAB_TAU} shows the simulation of $\sqrt{\text{MCRB}(\tau)}$ for different values of $L$, ranging from 1024 Samples ($\simeq 0.5\,\text{ms}$ with a symbol duration of $T=2^{-21}\,\text{s}$) up to 131072 Samples ($\simeq 62.5\,\text{ms}$). If we assume an $E_s/N_0$ of $20\,\text{dB}$ (which is realistic due to the high transmission power of DAB), a standard deviation of \mbox{$\sigma_{\tau} \approx 7.0\cdot 10^{-11}\,\text{s} $} (equivalent to \mbox{$2.1\,\text{cm}$}) is theoretically possible for \mbox{$L$ = 131072}.
\subsection{MCRB for Carrier Frequency Synchronization}
\begin{figure}[htbp]
\centerline{\includegraphics[width=0.5\textwidth]{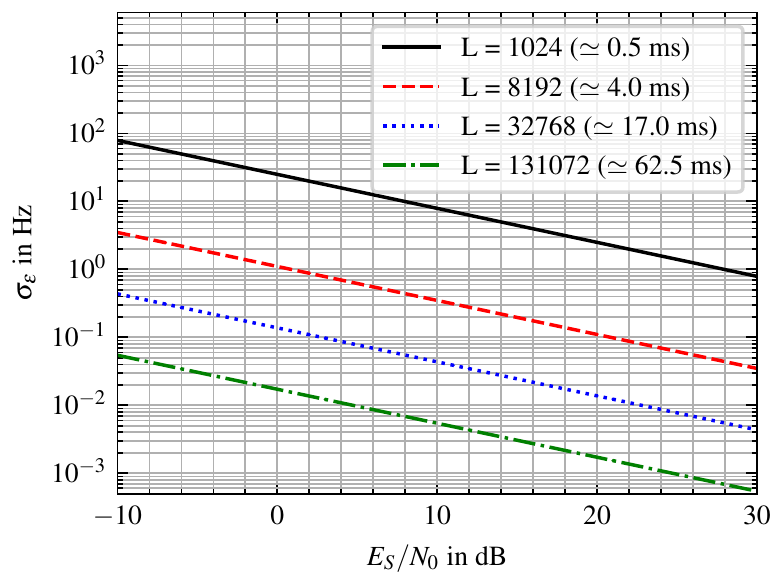}}
\caption{$\sqrt{\text{MCRB}(\epsilon)}$ for estimation of the carrier frequency offset as a function of the signal-to-noise ratio $E_S/N_0$ for different observation lengths $L$}
\label{DAB_CFO}
\end{figure}

In a second step, we will take a closer look at the MCRB for estimation of the carrier frequency offset $\epsilon$. With the same assumptions on the DAB waveform, it can be stated according to \cite{MCRB}:
  \begin{equation}
     \text{MCRB}(\epsilon) = \frac{3T}{2 \pi ^2 (LT)^3}\frac{N_0}{E_S}
\label{MCRB_CFO}
 \end{equation}
Contrary to $\text{MCRB}(\tau)$, the estimation variance now depends on the third power of the observation length $L$. A doubling of $L$ results in an increased estimation precision of factor 8. However, $\text{MCRB}(\epsilon)$ is completely independent of the used signal bandwidth $B$. Fig. \ref{DAB_CFO} shows a simulation of the $\sqrt{\text{MCRB}(\epsilon)}$ for different observation lengths. 
Choosing \mbox{$L$ = 131072} and assuming an $E_s/N_0$ of $20\, \text{dB}$ we can achieve a standard deviation of approx. \mbox{$2\cdot 10^{-3}\, \text{Hz}$.} 
 \\ In this section, the theoretical limits for synchronization accuracy were derived for DAB. It was shown that a time synchronization precision in the sub-meter range could be expected, fulfilling our requirements for indoor localization from section \ref{Concept}. Furthermore, the accuracy of CFO estimation is expected in the range of milli Hertz.

\section{Proposed Estimation Algorithm} \label{Framework}
\begin{figure}[htbp]
\centerline{\includegraphics[width=0.49\textwidth]{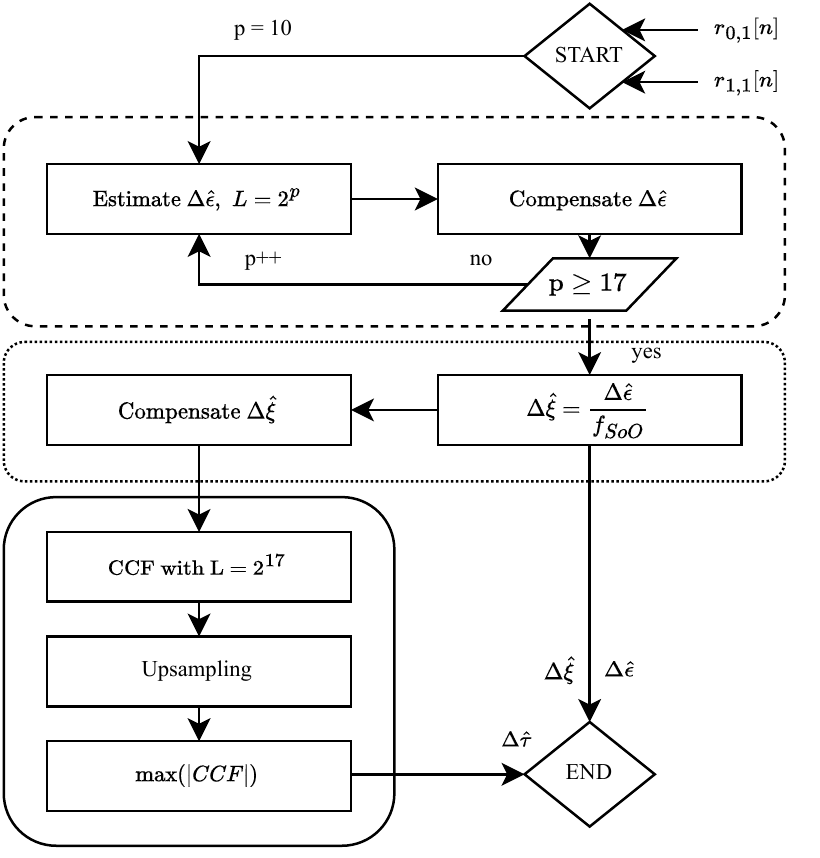}}
\caption{Process diagram of the proposed algorithm with iterative estimation of $\Delta \hat{\epsilon}$ (dashed box), calculation of $\Delta \hat{\xi}$ (dotted box) and estimation of the time synchronization $\Delta \hat{\tau}$ (solid box)}
\label{algo}
\end{figure}

In this section, we present an ultra-precise estimation algorithm for the differential synchronization parameter, initially stated in (\ref{const_tau}), (\ref{const_eps}), and (\ref{const_xi}). The proposed algorithm evaluates the cross-correlation function (CCF) between received SoO baseband signals, denoted as $R_{r_{i,j},r_{i,j}}$. Assuming we want to synchronize the base stations with indices 0 and 1, we can express the CCF with the baseband signals from (\ref{baseband}) as follows:
\begin{multline}
 R_{r_{0,1},r_{1,1}}[m] = \sum_{n = - \infty}^{\infty} r_{0,1}^*[n]\cdot r_{1,1}[n+m]  \\
    = \sum_{n = - \infty}^{\infty}  r_{0,1}^*(kT + \xi_{0,1} + \tau_{0,1})\cdot r_{1,1}(kT + m + \xi_{1,1}+\tau_{1,1})\\  \text{exp}(-j2\pi k T (\epsilon_{0,1} - \epsilon_{1,1})) + w_{R}[n]
\label{CCF}
\end{multline}
where $w_{R}[n]$ is a combined AWGN process considering the noise processes from both base stations and (.)* denotes the complex conjugate. Eq. (\ref{CCF}) shows that the exponent is a function of the differential carrier frequency offset \mbox{$\Delta \hat{\epsilon}$ = $\epsilon_{0,1}$ - $\epsilon_{1,1}$}. To estimate $\Delta \hat{\epsilon}$, we make use of the well-known relation between the phase of a signal and its instantaneous frequency:
\begin{equation}
    \Delta \hat{\epsilon} = -\frac{1}{2\pi}\frac{\partial \Delta\hat{\phi}}{\partial t}
\end{equation}
To create a derivable vector of phase estimates $\Delta\hat{\phi}$, we segment $r_{0,1}[n]$ and $r_{1,1}[n]$ into vectors of equal length $L= 2^p$, $p\in \mathbb{N}$. Afterwards, a CCF is computed pairwise between the segments of $r_{0,1}[n]$ and $r_{1,1}[n]$. The phase is extracted from the peak of the CCF according to:
\begin{equation}
    \Delta \hat{\phi} = \text{arg}(\text{max}(|R_{r_{0,1},r_{1,1}}|))
\end{equation}
A design parameter of the proposed estimation principle is the choice of the observation length $L$. The theoretical considerations from Section \ref{Theory} demand a large $L$ for high estimation precision. However, phase ambiguities are possible for large CFO values in combination with a large $L$, making the estimation infeasible. These ambiguities occur when $\Delta \hat{\phi} > 2\,\pi$ holds. The relationship between highest unambiguous resolvable $\Delta \hat{\epsilon}$ and $L$ is given by:
\begin{equation}
    \Delta \hat{\epsilon}_{max} = \frac{1}{TL}
\label{CFO_MAX}
\end{equation}
The presented estimation algorithm overcomes this issue by using an iterative approach for CFO estimation. 
\\ Fig. \ref{algo} shows the whole algorithm, including the iterative CFO estimation, calculation of the SCO, and the time synchronization within a process diagram.

The algorithm takes both received SoO baseband signals $r_{0,1}[n]$ and $r_{1,1}[n]$ as input and starts the iterative CFO estimation with a segment length of $L=2^{10}$ samples, allowing an initial CFO of $\Delta \hat{\epsilon}_{max} = 2048\,\text{Hz}$ (\ref{CFO_MAX}). After the estimation, $\Delta \hat{\epsilon}$ is compensated on $r_{0,1}[n]$ and $r_{1,1}[n]$. The remaining CFO is small enough to be estimated with a higher length $L$ without any phase ambiguities. This procedure is repeated until we reach $L=2^{17}$, which  - according to Fig. \ref{DAB_CFO} - already promises a good estimation precision in the range of mHz. After the last iteration, the differential CFO $\Delta \hat{\epsilon}$ is fully compensated. Next, we use the estimate of $\Delta \hat{\epsilon}$ to calculate the sampling clock offset $\Delta \hat{\xi}$  according to (\ref{sco}) with $f_{0,1}=f_{1,1}=f_{SoO}$, denoting the nominal frequency of the received SoO waveform. $\Delta \hat{\xi}$ is compensated on the signals by correcting the symbol durations with a Farrow structure, extensively explained in \cite{Farrow}. After a full compensation of $\Delta \hat{\epsilon}$ and $\Delta \hat{\xi}$, an estimate of the time synchronization parameter $\Delta \hat{\tau}$ is calculated by evaluating the CCF with $L=2^{17}$, promising accuracy in the range of centimeters according to Fig. \ref{MCRB_TAU}. To achieve this precision, upsampling is applied to the CCF. The estimated synchronization parameter  $\Delta \hat{\tau}$ is extracted at the up-sampled peak of the CCF. 

\section{Initial Measurement Results} \label{Results}
\begin{figure}[htbp]
\centerline{\includegraphics[width=0.5\textwidth]{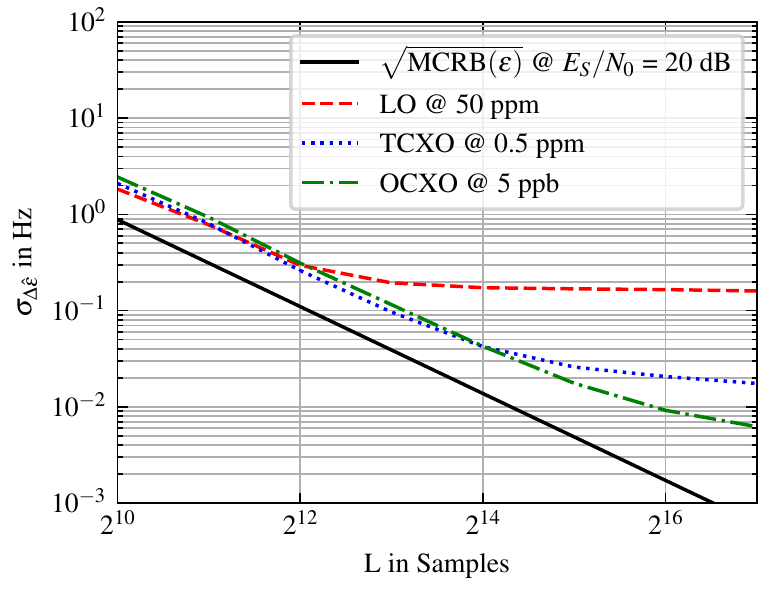}}
\caption{Comparison of the theoretical lower limit for carrier frequency offset estimation precision according to (\ref{MCRB_CFO}) with a fixed $E_S/N_0$ of $20\,\text{dB}$ to measured precision $\Delta \hat{\epsilon}$ as a function of the observation length $L$ for different oscillator types}
\label{meas_cfo}
\end{figure}

This section provides first measurements with the proposed system to verify the theoretical lower bounds from section \ref{Theory}. For carrying out the measurements, the testbed, formerly described in \cite{LPWAN}, was used in combination with SDRPlay Duo\footnote{https://www.sdrplay.com/rspduo/ (accessed May 2023)} frontends. For the measurements, the SoO waveform was received at a single antenna and fed into two frontends using a power splitter at a constant $E_S/N_0$ of $20\,\text{dB}$. This measurement method was chosen to obtain good comparability against the theoretical limits since the MCRB only considers white Gaussian noise. A setup with spatially separated stations could cause errors due to multipath propagation. To achieve reliable data points for calculating the standard deviations, each $\sigma$ is measured within one second of SoO data and averaged over N = 100 data sets. Furthermore, the whole evaluation process was conducted with three different oscillators. Since the performance of the whole algorithm depends on the principle of LO-sharing, it is worth looking at the influence of different oscillator types on the overall synchronization performance. The investigated oscillators, therefore, differ in their frequency stability and used technique. The internal oscillator of the SDRPlay Duo employs a temperature-compensated crystal oscillator (TCXO) with a frequency stability of $0.5\,\text{ppm}$. Besides the internal oscillator, a free running oscillator\footnote{https://www.ctscorp.com/wp-content/uploads/CA25C.pdf\newline (accessed May 2023)} (LO) with $50\,\text{ppm}$ and an oven controlled crystal oscillator\footnote{http://www.conwin.com/datasheets/cx/cx193.pdf (accessed May 2023)} (OCXO) with $5\,\text{ppb}$ are evaluated in the measurements.
Fig. \ref{meas_cfo} shows the comparison between theoretical limit and measurement for the carrier frequency offset $\Delta \hat{\epsilon}$ between two base stations as a function of the observation length $L$ for the different oscillator types. An important note at this point concerns the algorithm presented in the previous chapter. Since the proposed algorithm is calculating a CFO estimate from two consecutive phase estimates, we have to consider this in the comparison against the theoretical limit. Hence, the CFO estimation with two consecutive slices of length $L/2$ has to be compared to MCRB($\epsilon$) at observation length $L$. Fig. \ref{meas_cfo} already takes this fact into account.

From Fig. \ref{meas_cfo}, it can be seen that at least one oscillator nearly reaches the theoretical standard deviation according to $\sqrt{\text{MCRB}(\epsilon)}$. The OCXO features the best performance, which can be attributed to its good frequency stability. While the OCXO reaches the bound up to a minimum distance of $4.5\,\text{dB}$ at $L=2^{12}$ and a maximum CFO estimation precision of approx. $ 6\cdot \,10^{-3}\,\text{Hz}$ for $L=2^{17}$, the free running oscillator fails in this scenario with a more than one order of magnitude worse performance. This is probably caused by frequency fluctuations within the measurement duration of one second because the frequency is neither stabilized by a temperature compensation nor an oven. Furthermore, the curves for all oscillators are flattening out for high observation lengths. This is expected to be caused by the increasing influence of phase noise within large correlation lengths and minor temperature effects within the measurement interval, causing frequency instability. 
\\A similar evaluation can be conducted to check if the theoretical synchronization boundary $\text{MCRB}(\tau)$ can be reached in real-world measurements. This scenario is depicted in Fig. \ref{meas_synch}:

\begin{figure}[htbp]
\centerline{\includegraphics[width=0.5\textwidth]{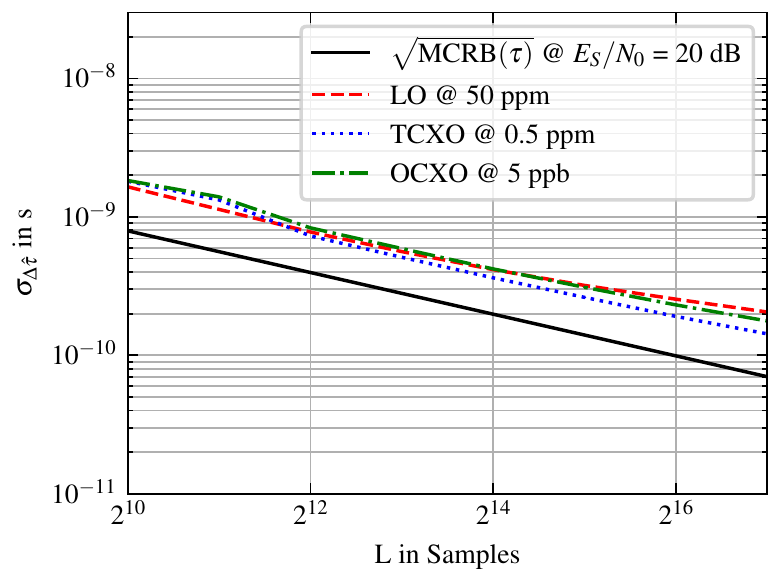}}
\caption{Comparison of theoretical lower limit for the time synchronization parameter $\tau$ according to (\ref{MCRB_TAU_CLOSED}) with a fixed $E_S/N_0$ of $20\,\text{dB}$ to measured estimation synchronization precision $\Delta \hat{\tau}$ as a function of the observation length $L$ for different oscillator types}
\label{meas_synch}
\end{figure}

The measurement results in Fig. \ref{meas_synch} suggest that the influence of different oscillator types on the synchronization accuracy is not as significant as in Fig. \ref{meas_cfo}. All oscillator types nearly feature the same precision and reach $\sqrt{\text{MCRB}(\tau)}$ up to a minimum difference of $\approx 3.2\,\text{dB}$ for $L=2^{13}$ with a maximum synchronization accuracy of approx. $2\cdot10^{-10}\,\text{s}$  (equivalent to $6\,\text{cm}$) at $L=2^{17}$.

\section{Summary and Conclusion} \label{Conclusion}
This article presented an approach for wireless synchronization of spatially distributed base stations based on a Signal of Opportunity. The proposed software-based synchronization concept uses base station frontends with multiple channels sharing an oscillator, which makes it easily adaptable to many state-of-the-art frontends. The theoretical limits according to the Modified Cramer-Rao Bound were derived, where it was shown that a sufficiently high $E_S/N_0$ of $20\,\text{dB}$ enables a time synchronization accuracy of $7.0\cdot10^{-11}\,\text{s}$ (equivalent to $\,2.1\,\text{cm}$) and a carrier frequency offset estimation accuracy of $2\cdot10^{-3}\,\text{Hz}$. An estimation algorithm was presented that evaluates the cross-correlation function between two base stations to find estimates of the synchronization parameters. In real-world measurements, the introduced algorithm was tested against the theoretical limits. A  synchronization accuracy of approx. $2\cdot10^{-10}\,\text{s}$ (equivalent to 6\,\text{cm}) was achieved, which reaches the theoretical limit up to a minimum distance of $3.2\,\text{dB}$. The measured accuracy of the carrier frequency estimation was found at approx. $6\cdot 10^{-3}\,\text{Hz}$, only $4.5\,\text{dB}$ apart from its theoretical limit.
\\In future work, extending the system to different Signals of Opportunity can increase the achievable accuracy. Further investigations of the influence of phase noise on the synchronization performance are necessary. Finally, methods can be developed that allow a reduction of the amount of data that is necessary to perform the synchronization. Document \cite{SWITCH} proposes an adaptation of the presented concept to enable the use of a single receive channel by switching the carrier frequency, dramatically reducing the necessary amount of data and computational load.

\section*{Acknowledgement}
This work is part of the research project 5G-Flexi-Cell (grant no. 01MC22004B) funded by the German Federal Ministry for Economic Affairs and Climate Action (BMWK) based on a decision taken by the German Bundestag.



\begin{thebibliography}{00}
\bibitem{GPS_Indoor}
S. Sadowski and P. Spachos, “RSSI-Based Indoor Localization With the Internet of Things,” IEEE Access, vol. 6, pp. 30149–30161, 2018.

\bibitem{IndoorLocOverview}
A. Billa, I. Shayea, A. Alhammadi, Q. Abdullah, and M. Roslee, “An Overview of Indoor Localization Technologies: Toward IoT Navigation Services,” in 2020 IEEE 5th International Symposium on Telecommunication Technologies (ISTT), Nov. 2020, pp. 76–81.

\bibitem{UWBTDoA}
Y. Cheng and T. Zhou, “UWB Indoor Positioning Algorithm Based on TDOA Technology,” in 2019 10th International Conference on Information Technology in Medicine and Education (ITME), Aug. 2019, pp. 777–782. 



\bibitem{Wired_Wireless_Comparison}
S. Leugner, M. Pelka, and H. Hellbrück, “Comparison of wired and wireless synchronization with clock drift compensation suited for U-TDoA localization,” in 2016 13th Workshop on Positioning, Navigation and Communications (WPNC), Oct. 2016, pp. 1–4.

\bibitem{PTP}
A. Mahmood, R. Exel, and T. Sauter, “Delay and Jitter Characterization for Software-Based Clock Synchronization Over WLAN Using PTP,” IEEE Transactions on Industrial Informatics, vol. 10, no. 2, pp. 1198–1206, May 2014.





\bibitem{Troeger}
H.-M. Tröger, J. Robert, L. Patino-Studencki, and A. Heuberger, “A Comparison of Opportunistic Signals for Wireless Syntonization Using the Modified Cramér–Rao Lower Bound,” NAVIGATION, vol. 64, no. 3, pp. 351–363, 2017.

\bibitem{PN}
A. Chorti and M. Brookes, “A Spectral Model for RF Oscillators With Power-Law Phase Noise,” IEEE Transactions on Circuits and Systems I: Regular Papers, vol. 53, no. 9, pp. 1989–1999, Sep. 2006.

\bibitem{MCRB}
A. N. D’Andrea, U. Mengali, and R. Reggiannini, “The modified Cramer-Rao bound and its application to synchronization problems,” IEEE Transactions on Communications, vol. 42, no. 234, pp. 1391–1399, Feb. 1994.

\bibitem{DAB_SPECTRUM}
J. Robert, “Digital Audio Broadcasting (DAB),” in Wiley Encyclopedia of Electrical and Electronics Engineering, John Wiley \& Sons, Ltd, 2021, pp. 1–12.

\bibitem{Farrow}
C. W. Farrow, “A continuously variable digital delay element,” in 1988., IEEE International Symposium on Circuits and Systems, Jun. 1988, pp. 2641–2645 vol.3

\bibitem{LPWAN}
M. Michael, J. Robert, C. Neumüller, and A. Heuberger, “IoT Cloud RAN Testbed for Indoor Localization based on LPWANs,” in 2021 8th International Conference on Internet of Things: Systems, Management and Security (IOTSMS), Dec. 2021, pp. 1–6. 

\bibitem{SWITCH}
S. Klob, T. Maul, J. Robert, “Low-Cost and Ultra-Precise Synchronization
Concept for TDoA Localization of Dairy Cows”, 2023, submitted for publication at IPIN 2023.
\end{thebibliography}
\end{document}